\documentclass[a4paper,11pt,twoside]{article}
\usepackage[paperheight=29.7cm,paperwidth=21cm,outer=2.25cm,inner=2.25cm,top=2cm,bottom=2cm,headheight=13.6pt,includehead,includefoot]{geometry} 
\usepackage{silence}
\newcommand{\textlineskip}{\baselineskip=13pt}

\providecommand{\keywords}[1]{\noindent\textbf{Keywords:} #1}

\def\fnm#1{$^{\mbox{\scriptsize #1}}$}
\def\fnt#1#2{\footnotetext{\kern-.3em%
          {$^{\mbox{\scriptsize #1}}$}{#2}}}
          
\newcommand{\fcaption}[1]{\caption{#1}}
\newcommand{\tcaption}[1]{\caption{#1}}

\newtheorem{definition}{Definition}

\usepackage{fancyhdr}

\fancypagestyle{pagestyle}{%
  \fancyhf{}
  \fancyhead[OR]{A. Bonomi, T. De Min, E. Zardini, E. Blanzieri, V. Cavecchia, and D. Pastorello}
  \fancyhead[EL]{Quantum Annealing Learning Search Implementations}
  \fancyfoot[C]{\thepage}%
}

\fancypagestyle{firstpagestyle}{%
  \fancyhf{}
  
  \fancyfoot[C]{\thepage}
}

\renewcommand{\thefootnote}{\fnsymbol{footnote}}  

\usepackage{amsmath}
\usepackage{amssymb}
\usepackage[linesnumbered]{algorithm2e}

\usepackage[font=footnotesize]{caption}
\usepackage{graphicx}

\usepackage{bm}
\bmdefine{\bz}{z}
\bmdefine{\bTheta}{\Theta}

\def\si{\sigma}

\def\cA{{\ca A}}

\def\vz{{\bf z}}

\def\beq{\begin{eqnarray}}
\def\eeq{\end{eqnarray}}

\newcommand{\ca}[1]{{\cal #1}}         

\def\hs{\hspace{0.3cm}}

\let\oldparagraph\paragraph
\renewcommand{\paragraph}[1]{\oldparagraph{#1}\mbox{}\\}

\usepackage{multirow}

\usepackage[hyphens]{url}
\usepackage[hyperfootnotes=false]{hyperref}
\usepackage{xcolor}
\hypersetup{
  colorlinks,
  linkcolor={black},
  citecolor={black},
  urlcolor={black}
}
\usepackage[noabbrev,capitalise]{cleveref}

\usepackage{soul}

\begin{document}

\normalsize\textlineskip
\thispagestyle{firstpagestyle}
\setcounter{page}{1}

\centerline{\Large\bf Quantum Annealing Learning Search} 
\vspace*{0.035truein}
\centerline{\Large\bf Implementations}

\vspace{25pt}

\begin{center}
    \textbf{Andrea Bonomi$^\star$}, \hspace{1pt}
    \textbf{Thomas De Min$^\star$}, \hspace{1pt}
    \textbf{Enrico Zardini$^{\star \dagger}$}, \\  
    \textbf{Enrico Blanzieri$^{\star \ddagger}$}, \hspace{1pt}
    \textbf{Valter Cavecchia$^{\mathsection}$}, \hspace{1pt}
    \textbf{Davide Pastorello$^{\star \ddagger}$}

    \vspace*{10pt}

    $^\star$ Department of Information Engineering and Computer Science\\ University of Trento \\ 
    $ $ via Sommarive 9, 38123 Povo, Trento, Italy

    \vspace*{10pt}

    $^\dagger$ enrico.zardini@unitn.it

    \vspace*{10pt}

    $^\ddagger$ Trento Institute for Fundamental Physics and Applications \\ 
    $ $ via Sommarive 14, 38123 Povo, Trento, Italy

    \vspace*{10pt}

    $^\mathsection$ Institute of Materials for Electronics and Magnetism (CNR) \\
    $ $ via alla Cascata 56/c, 38123 Povo, Trento, Italy
    
\end{center}

\vspace*{15pt}

\begin{abstract}
\noindent
This paper presents the details and testing of two implementations (in C++ and Python) of the hybrid quantum-classical algorithm Quantum Annealing Learning Search (QALS) on a D-Wave quantum annealer. QALS was proposed in 2019 as a novel technique to solve general QUBO problems that cannot be directly represented into the hardware architecture of a D-Wave machine. Repeated calls to the quantum machine within a classical iterative structure and a related convergence proof originate a learning mechanism to find an encoding of a given problem into the quantum architecture. The present work considers the Number Partitioning Problem (NPP) and the Travelling Salesman Problem (TSP) for the testing of QALS. The results turn out to be quite unexpected, with QALS not being able to perform as well as the other considered methods, especially in NPP, where classical methods outperform quantum annealing in general. Nevertheless, looking at the TSP tests, QALS has fulfilled its primary goal, i.e., processing QUBO problems not directly mappable to the QPU topology.

\vspace*{10pt}
\end{abstract}

\setcounter{footnote}{0}
\renewcommand{\thefootnote}{\alph{footnote}}

\vspace*{1pt}\textlineskip

\section{Introduction}
\label{sec:intro}
\noindent
Quantum annealing \cite{PhysRevE.58.5355} is an optimization technique based on the physical property of quantum systems of reaching with high probability the global energy minimum even when the energy profile is complex and rich of local minima. When the energy profile is set to be a representation of a computational problem of interest, the measurement outcome represents the solution of the problem. The approach has been popularised by the Burnaby-based company D-Wave, which builds quantum annealers and puts them on the market of hardware and computational services. The D-Wave systems exploits the physical realisation of an Ising Model to solve Quadratic Unconstrained Binary Optimisation (QUBO) problems. The topology of the physically-realised Ising model is not complete and different layouts (Chimera and Pegasus) have been proposed in the current versions of the D-Wave machines. Notably, these commercial machines reach a number of qubits in the order of thousands (2048 for Chimera and 5700 for Pegasus), far bigger than general-purpose machines, which have currently a number of qubits in the order of hundreds. This led to the expectation that the application of quantum annealers could in principle, for specific problems, provide solutions that outperform classical counterparts with a growing scientific literature of promising applications (see \cite{bian2020solving} for a recent example).

In general, the practical application of quantum computing leads naturally to the proposition of Hybrid Quantum Classical approaches. On the one hand, the motivation is the relative immaturity of quantum computing architectures in the so-called NISQ era \cite{preskill2018quantum} and hybrid approaches can effectively overcome some limitations of the current quantum computing machines. On the other hand, it is reasonable to assume that the interaction between classical and quantum parts of an algorithm can better exploit the advantages of both paradigms. In particular, hybrid quantum-classical approaches have been proposed for both general-purpose quantum machines \cite{mcclean2016theory}, and for quantum annealing architectures \cite{ayanzadeh2020reinforcement}.

The paradigm of hybrid quantum-classical computing was applied by Pastorello and Blanzieri \cite{QALS}, who proposed a schema called Quantum Annealing Learning Search (QALS) based on: i) a classical external annealing loop and ii) the inclusion of a tabu mechanism in the energy profile in order to overcome the limitation in terms of topology of the current quantum annealers. A complete-topology QUBO is iteratively and randomly mapped to the available physical topology and the tabu information is additively integrated into the energy profile. The authors, by reducing to a result on general tabu search, presented an argument for the asymptotic convergence of the external annealing loop (i). However, the work did not include an empirical evaluation of QALS. More recently, and in the more general setting of adiabatic quantum computing, an analogous schema was produced \cite{pastorello_learning_2021} and the asymptotic convergence empirically verified with a simulation. Still, the tabu-based schema (ii) was not applied to real problems on a real quantum computing architecture, consideration that motivates the present work.

In this paper, we present two implementations of QALS for the quantum annealer and test them on two different problems. The first implementation is in C++ with the purpose of increasing the performance of the classical part, the second one is in Python in order to have a smooth interface with the D-Wave machine we used. The implementations were tested on the Number Partitioning Problem (NPP) that has a direct QUBO representation \cite{QUBOsurvey} and on the Travelling Salesman Problem (TSP) that requires to represent also penalties in the corresponding QUBO formulation. Results and performances are compared against classical competitors and the hybrid solution provided by D-Wave. The results show that the proposed method permits to tackle bigger problems w.r.t. the standard hybrid solution.

The paper is organised as follows. Sections \ref{sec:background} and \ref{subsec:qals} present the background and QALS, respectively. Section \ref{sec:implementations} describes the implementations and Section \ref{sec:experimental-results} discusses NPP, TSP, and the results. Finally, we draw some conclusions in Section \ref{sec:conclusion}.

\section{Background}
\label{sec:background}
\noindent
Quantum annealing (QA) is a type of heuristic search used to solve optimization problems~\cite{PhysRevE.58.5355}. The solution of a given problem corresponds to the so-called \emph{ground state} of a quantum system with total energy described by a \emph{problem Hamiltonian} $H$ on the Hilbert space where the considered quantum system is described. The annealing procedure is implemented by a time evolution of the quantum system towards the \emph{ground state} (i.e. the less energetic physical state) of the problem Hamiltonian. 

QA can be physically realized considering a quantum spin glass that is a network of qubits arranged on the vertices of a graph $\langle V,E\rangle$, with $|V|=n$, whose edges $E$ represent the couplings among the qubits. The total energy of such a system is represented by the Hamiltonian $H(\boldsymbol\Theta)$ of the quantum spin glass,
\begin{equation}\label{HP}
H(\boldsymbol\Theta)=\sum_{i\in V} \theta _i \sigma_z^{(i)} +\sum_{(i,j)\in E} \theta_{ij}\sigma_z^{(i)} \sigma_z^{(j)}.
\end{equation}
$H(\boldsymbol\Theta)$ is an operator on the $n$-qubit Hilbert space $\mathsf H=({\mathbb C}^2)^{\otimes n}$ where $\sigma_z^{(i)}$ acts as the Pauli matrix
\[
\sigma_z=\begin{bmatrix}
1 & 0\\
0 & -1
\end{bmatrix}
\]
on the $i$th tensor factor and as the $2\times 2$ identity matrix on the other tensor factors. 
The coefficient matrix $\boldsymbol\Theta$ is the symmetric square $n\times n$ matrix of the real coefficients of $\mathsf E$, called \emph{weights}, defined as
\begin{equation}\label{w}
\boldsymbol\Theta_{ij}:=\left\{
\begin{array}{ll}
\theta_i, & i=j,\\
\theta_{ij}, & (i,j)\in E,\\
0, & (i,j)\not\in E.
\end{array}\right.
\end{equation}
The coefficient $\theta_i$ physically corresponds to the local field on the $i$th qubit and $\theta_{ij}$ to the coupling between the qubits $i$ and $j$.  

The Pauli matrix $\sigma_z$ has two eigenvalues corresponding to the binary states of the qubit, $\{-1,1\}$, and thus, the system~\eqref{HP} has the spectrum of eigenvalues corresponding to all possible values of the cost function given by the energy of the well-known \emph{Ising model}:
\begin{equation}\label{costf}
    \mathsf E(\boldsymbol{\Theta}, \boldsymbol z)=\sum_{i\in V} \theta_i z_i +\sum_{(i,j)\in E} \theta_{ij}z_i z_j,\quad \boldsymbol z=(z_1,...,z_n)\in\{-1,1\}^{|V|}.
\end{equation}
In practice, the annealing procedure drives the system into the ground state of $H(\Theta)$ corresponding to the spin configuration encoding the solution:
\begin{equation}\label{argmin_zE}
\boldsymbol z^*=\arg\!\!\!\!\!\!\!\!\min_{\boldsymbol z\in\{-1,1\}^{|V|}} \mathsf E(\boldsymbol\Theta,\boldsymbol z).
\end{equation}
Given a problem, the annealer is initialized by a suitable choice of the weights $\boldsymbol\Theta$ and the binary variables $z_i\in\{-1,1\}$ are physically realized by the outcomes of measurements on the qubits located in the vertices $V$. In order to solve a general optimization problem by QA, one needs to obtain the correct \textit{encoding} of the objective function in terms of the cost function~\eqref{costf}, which is hard in general. 

\subsection{QUBO model}
\label{subsec:QUBO}
\noindent
Quadratic Unconstrained Binary Optimization (QUBO) problems are NP-hard problems traditionally used in computer science. A general QUBO problem is defined by the minimization of a function $f:\{0,1\}^n\rightarrow \mathbb R$ of the form:
    \begin{equation}
        f(x) = \displaystyle\sum_{i=1}^n Q_{ii}x_i + \sum_{i<j}Q_{ij}x_i x_j,
        \label{eq:function_QUBO}
    \end{equation}
where $Q_i$ and $Q_{ij}$ are real coefficients that can be arranged into an $n \times n$ upper-triangular real matrix $Q$. Therefore, a QUBO problem can be represented as:
    \begin{equation}
        \displaystyle\min_{x\in\{0,1\}^n} x^T Q x.
        \label{minQUBO}
    \end{equation}
The QUBO model (\ref{minQUBO})  covers a remarkable range of applications in combinatorial optimization: optimization problems on graphs, facility location problems, resource allocation problems, clustering problems, set partitioning problems, various forms of assignment problems, sequencing and ordering problems \cite{QUBOsurvey}.

\subsection{D-Wave Pegasus topology}
\label{subsubsec:pegasus}
\noindent
In order to apply quantum annealing, it is necessary to embed (minor embed, in the D-Wave glossary) the QUBO matrix in the topology of a Quantum Processing Unit (QPU) such as D-Wave Pegasus. Basically, the problem variables must be mapped to QPU qubits; further details can be found here \cite{minor_embedding, minor_embedding_example}.

\begin{figure}
    \centering
    \includegraphics[width=.65\linewidth]{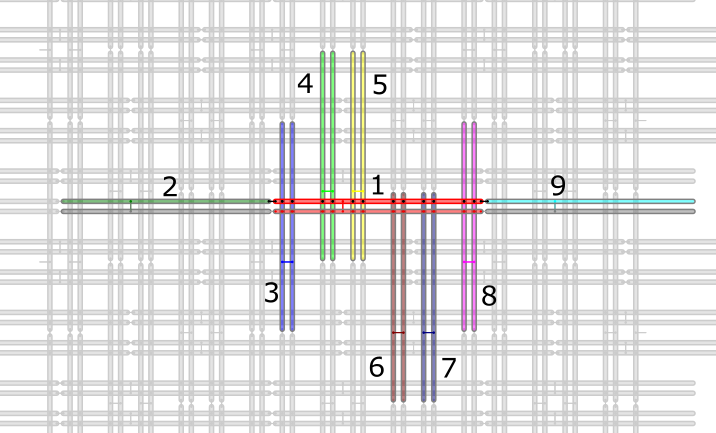}
    \vspace{13pt}
    \fcaption{Pegasus topology  \href{https://docs.dwavesys.com/docs/latest/c\_gs\_4.html}{(source link)}.}
    \label{fig:pegasustopology}
\end{figure}

Pegasus is the successor of Chimera and presents 5700 qubits. In this topology, qubits are oriented vertically or horizontally and the similarly aligned qubits are also shifted (look at \cref{fig:pegasustopology}). In detail, there are three types of couplers \cite{pegasus}:
\begin{itemize}
    \item internal couplers: connect pairs of orthogonal qubits (each qubit is linked to 12 other qubits by means of internal couplers);
    \item external couplers: connect vertical (horizontal) qubits to vertical (horizontal) adjacent qubits;
    \item odd couplers: connect similarly aligned pairs of qubits.
\end{itemize}

\subsection{Embedding Composite} 
D-Wave provides several tools for pre and post-processing to be used in conjunction with samplers. Among them, there is the tool named \textit{Embedding Composite} \cite{embedding_composite}, which automatically minor-embeds the given problem into the topology of the chosen QPU. In particular, a new minor-embedding is calculated each time the sampling method is called.

\subsection{D-Wave Hybrid}
\label{subsubsec:hybrid}
\noindent
As an alternative to simulated and quantum annealing, D-Wave provides the so-called \textit{Hybrid} annealing, a framework that allows running multiple solvers in parallel, either classical or not. In particular, a branch could be represented by a classical technique such as a Tabu Search or by a workflow of the form \textit{decomposer} - \textit{sampler} - \textit{composer}. As concerns the latter, the decomposer splits the given problem into subproblems, which are solved by the sampler and whose local solutions are recomposed by the composer. The sampler could be simulated annealing, quantum annealing, but also another classical method or something more complex (such as parallel solvers). In this work, the default hybrid configuration provided by D-Wave has been used; details about the Python class used and the default solver parameters/properties are available here \cite{leap_hybrid_sampler, solver_params, solver_properties}.

\section{Quantum Annealing Learning Search (QALS)}
\label{subsec:qals}

\noindent
Quantum Annealing Learning Search (QALS) is a guided meta-heuristic approach designed specifically to solve optimization problems that cannot be directly represented into the architecture of a quantum annealer. The idea of the approach  suggested in~\cite{QALS} and further generalized in~\cite{pastorello_learning_2021} is to initialize the QA by adding the so-called \emph{tabu matrix} $S$ to the weight matrix $\boldsymbol\Theta$. The contribution of ${S}$ is to penalize already visited solutions preventing a redundant search in the solution space. Let us assume we have a set of $k$ solutions $\{\boldsymbol z_j\}_{j=1,...,k}$ to penalize. The matrix $\boldsymbol S$ is constructed as the sum of outer products of the solutions to be penalized:
\begin{equation}\label{matrix}
    S=\sum_{j=1}^k (\boldsymbol z_j \boldsymbol z_j^T -\boldsymbol I+\mathrm{diag}\, \boldsymbol z_j),
\end{equation}
where $\boldsymbol I$ is the identity matrix of size $n$, and $\mathrm{diag}\boldsymbol z_j$ is a diagonal matrix from the vector $\boldsymbol z_j$. By construction, the tabu matrix $\boldsymbol{S}$ introduces \emph{energetic penalties} on the solutions $\{\boldsymbol z_j\}_{j=1,...,k}$ in the spectrum of the Hamiltonian $H(\boldsymbol\Theta+ S)$, which is the energy function $\boldsymbol z \mapsto \mathsf E(\Theta+ S,\boldsymbol z)$.

\begin{algorithm}[p]
\footnotesize
\KwData{Matrix $Q$ of order $n$ encoding a QUBO problem,  annealer adjacency matrix $\cA$ of order $n$}
\KwIn{Energy function of the annealer $\mathsf E(\bTheta, \vz)$, permutation modification function $g(P,p)$, solution modification function $h(\vz)$, minimum probability $0<p_\delta<0.5$ of permutation modification, probability decreasing rate $\eta>0$, candidate perturbation probability $q>0$, number $N$ of iterations at constant $p$, initial balancing factor $\lambda_0>0$, number of annealer runs $k\geq 1$, termination parameters $i_{max}$, $N_{max}$, $d_{min}$ }
\KwResult{$\vz^*$ vector with $n$ elements in $\{-1,1\}$ solution of the QUBO problem} 
\SetKwProg{ffun}{function}{:}{}
\ffun{$f_Q$($x$)}{     \KwRet 
 $x^TQx$ \;}
 $P \gets I_n$\;
 $p \gets 1$\;
 $P_1 \gets g(P,1); P_2 \gets g(P,1)$; \tcp*[h]{\it generate two permutation matrices perturbing the identity}\\
 $\bTheta_1 \gets P_1^TQP_1\circ \cA$;  $\bTheta_2 \gets P_2^TQP_2\circ \cA$; \tcp*[h]{\it weights initialization}\\
 run the annealer $k$ times with weights $\bTheta_1$ and $\bTheta_2$\\
 $\vz_1 \gets P_1^T \widehat{\mbox{argmin}}_\vz(\mathsf E(\bTheta_1,\vz))$;  $\vz_2 \gets P_2^T \widehat{\mbox{argmin}}_\vz(\mathsf E(\bTheta_2,\vz))$; \tcp*[h]{\it estimate energy argmin, $P_1^T$ and $P_2^T$ map back the variables}\\
$ f_1 \gets f_Q(\vz_1); f_2 \gets f_Q(\vz_2)$ ; \tcp*[h]{\it evaluate $f_Q$}\\
\tcp*[h]{\it use the best to initialize $\vz^*$ and $P^*$; use the worst to initialize $\vz'$ }\\
  \eIf{$f_1<f_2$}{$\vz^* \gets \vz_1$; $f^* \gets f_1$; $P^* \gets P_1$ $\vz' \gets \vz_2$\;}{$\vz^* \gets \vz_2$ ; $f^* \gets f_2;$ $P^* \gets P_2$; $\vz' \gets \vz_1$\;}
\lIf( \tcp*[h]{\it use $\vz'$ to initialize the tabu matrix $S$}){$f_1\not =f_2$}{$S \gets \vz' \otimes \vz'-I_n+diag(\vz')$}
\lElse( \tcp*[h]{\it otherwise set all the elements of $S$ to zero}){$S \gets 0$}
$e \gets 0$; $d \gets 0$; $i \gets 0$; $\lambda\gets\lambda_0$\;
\Repeat{$i=i_{max}$ or ($e+d \geq N_{max}$ and $d<d_{min}$) }{
$Q' \gets Q+\lambda S$; \tcp*[h]{\it scale and add the tabu matrix   } \\ 
\If{$N$ divides $i$}{$p \gets p-(p-p_\delta)\eta$;}
$P \gets g(P^*,p)$; \tcp*[h]{\it modify permutation $P^*$}\\
$\bTheta' \gets P^TQ'P\circ \cA$; \tcp*[h]{\it weights initialization }\\
run the annealer $k$ times with weights $\bTheta'$\\ 
$\vz' \gets P^T \widehat{\mbox{argmin}}_\vz(\mathsf E(\bTheta',\vz))$; \tcp*[h]{\it estimate energy argmin, $P^T$ maps back the variables}\\
with probability $q\hs$ $\vz' \gets h(\vz',p)$; \tcp*[h]{\it possibly perturb the candidate}\\  
 \eIf{$\vz' \not= \vz^*$} {
$f' \gets f_Q(\vz')$; \tcp*[h]{\it evaluate $f_Q$}\\ 
 \eIf{$f'<f^*$}{$swap(\vz',\vz^*)$; $f^* \gets f'$; $P^* \gets P$; $e \gets 0$; $d \gets 0$; \tcp*[h]{\it $\vz'$ is better}\\ 
 $S \gets S+\vz' \otimes \vz'-I_n+diag(\vz')$; \tcp*[h]{\it use $\vz'$ to update the tabu matrix $S$}}
 { $d \gets d+1$\;
  with probability $(p-p_\delta)^{(f'-f^*)}$ $swap(\vz',\vz^*)$; $f^* \gets f'$; $P^* \gets P$; $e \gets 0$ \; 
 }
   update the balancing factor $\lambda$ with $\lambda\leq\lambda_0$\;
 }
 {$e \gets e+1$\;}
 $i \gets i+1$\;
 }
\Return $\vz^*$\;
\vspace{10pt}
\caption{Quantum Annealing Learning Search for QUBO problems.}
\label{Alg}
\end{algorithm}

The QALS scheme is based on an iterative procedure of candidate solutions generation by QA and probabilistic acceptance/rejection. The rejected solutions are used to update the tabu matrix ${S}$ and new candidate solutions are generated perturbing the weights as described below. Suboptimal acceptance is allowed and the perturbation of weights is proportional to a decreasing temperature parameter; so, the classical part of the hybrid algorithm presents a simulated annealing (SA) structure. 

The tabu strategy and the evolving representation of a general QUBO problem of QALS into the annealer can be summarized as follows: the matrix $Q$ representing the objective quadratic form $f_Q$ is piecewise mapped into the annealer architecture in the sense that only some elements of the QUBO matrix are selected by $P$ at each iteration; at the same time, $Q$ is deformed by means of the tabu matrix to energetically penalize the spin configurations corresponding to the solutions that are far from the optimum. More precisely, let us define the mapping $\mu$ that is employed to solve QUBO problems:
\beq\label{encod0}
\mu[f_Q]:= \mathsf E(P^T Q P\circ \cA ,P\vz),
\eeq 
where $P$ is a permutation matrix of order $n$ and $P^T$ its transpose, $\cA$ is the adjacency matrix of the graph describing the hardware topology of the quantum annealer and $\circ$ is the Hadamard product. Thus, the action of $\mu$ is realized by mapping some elements of $Q$, selected by $P$, into the weights. The tabu-implementing encoding of $Q$ into the annealer induced by the permutation matrix $P$ turns out to be:
\beq\label{encod}
\,\mu[f_Q](\vz)=\mathsf E( P^T (Q+\lambda S) P \circ \cA, P\vz),
\eeq 
where the scaling factor $\lambda$ regulates the contribution of the tabu matrix $S$, which is initialized to zero. 

The QALS scheme to solve general QUBO problems is presented in Algorithm \ref{Alg}. The tabu-implementing encodings are generated with permutations according to formulas (\ref{encod0}) and (\ref{encod}). To this end, we call an additional function $P\mapsto g(P,p)$ that modifies a permutation by considering for shuffling each element with probability $p$. In practice, the function $g$ produces the permutations that induce the encodings of the objective function into the annealer architecture (Algorithm~\ref{Alg}, lines 5 and 23). The mappings of the problem parameters to the annealer weights take into account the actual annealer topology represented by the graph matrix $\cA$ (Algorithm~\ref{Alg}, lines 6 and 24). The permutations are also used to remap the solutions found by the annealer to the initial space of solutions of the problem (Algorithm~\ref{Alg}, lines 8 and 26) and to represent the best map $\mu^*$.

As concerns the action of  $g$, if $p=1$, the resulting permutation is purely random; with $0<p<1$, the permutation resembles partially the initial one; if $p=0$, the output permutation would coincide with the input one. However, this last circumstance does not occur because the probability of an element to be shuffled decreases periodically to the value $0<p_\delta<0.5$ with rate $\eta$ (Algorithm~\ref{Alg}, lines 20-22). 

The matrix $Q$ of the QUBO problem interacts additively with the tabu matrix $S$ scaled by a balancing factor $\lambda$ (Algorithm~\ref{Alg}, line 19) in order to guide the search of the solution by quantum annealing with an energy profile consistent with (\ref{encod}). A crucial effect of this summation is that, in its iterative part, the algorithm does not search anymore just for solutions of subproblems as done instead in the initialization phase. In fact, $S$ contains information about the bad candidates (Algorithm~\ref{Alg}, lines 15 and 32) whose objective function values are greater than $f^*$. 
Moreover, the balancing factor $\lambda$, initially set to $\lambda_0$, is decreased during the search (Algorithm~\ref{Alg}, line 37); the goal of $\lambda$ is to avoid the action of the tabu matrix $S$ obscuring the information about $f_Q$ carried by $Q$. In general, $\lambda$ is a decreasing function of the number of bad candidates penalized by $S$.
The candidate solution found by the annealer (Algorithm~\ref{Alg}, line 26) is then perturbed with probability $q$ (Algorithm~\ref{Alg}, line 27) by the function $h(\vz', p)$ (this function flips any entry of $\vz'$ with probability $p$) in order to guarantee the convergence \cite{QALS}.

At line 35 of Algorithm~\ref{Alg}, a suboptimal solution is accepted with probability $(p-p_\delta)^{(f'-f^*)}$. By direct comparison with the common rule of acceptation in simulated annealing, namely $(p-p_\delta)^{(f'-f^*)}=e^{-\frac{(f'-f^*)}{T}}$, it is possible to provide an interpretation of $p$ by observing that $(p-p_\delta)=e^{-1/T}$. Namely, in terms of simulated annealing, the parameter $p$ is related to the temperature by $T=-{\ln}^{-1}[p-p_\delta]$, so $T\rightarrow 0$ as $p\rightarrow p_\delta$.

The cycle of Algorithm~\ref{Alg} ends due to convergence or when the maximum number of iterations is reached.   
Line 17 of Algorithm~\ref{Alg} defines three counters for controlling the end of the cycle: $e$ counts the number of consecutive times that the current best solution is generated (Algorithm~\ref{Alg}, line 39); $d$ counts the number of times the current best solution and the new solution differ and the current one is better (Algorithm~\ref{Alg}, line 34); finally, the variable $i$ simply counts the number of iterations. These counters are compared against input parameters in the termination condition (Algorithm \ref{Alg}, line 42).

\section{Implementations}
\label{sec:implementations}
\noindent
We developed two implementations of QALS, one in C++ and the other in Python. Both required systematic testing and some optimizations with respect to the original, mainly mathematical, formulation of QALS (the pseudocode, valid for both of them, is available in Algorithm~\ref{algo:main}). This section presents such general improvements and the details of each implementation.

\begin{algorithm}[p]
\footnotesize
\KwData{Matrix $Q$ of order $n$ encoding a QUBO problem, annealer adjacency matrix $\cA$ of order $n$}
\KwIn{\label{algo:main:inputs}Energy function of the annealer $\mathsf E(\bTheta, \bz)$, permutation modification function $g(P,p)$, minimum probability $p_\delta$ of permutation modification, probability decreasing rate $\eta$, tabu inversion probability $q$, number $N$ of iterations at constant $p$, scaling factor $\lambda_0$, number of annealer runs $k$, termination parameters $i_{max}$, $N_{max}$, $d_{min}$ }
\KwResult{$\bz^*$ vector with $n$ elements in $\{0,1\}$ solution of the QUBO problem} 
\SetKwProg{ffun}{function}{:}{}
\ffun{$f_Q$($Q$,$x$)}{     \KwRet 
 $x^TQx$ \;}
 $m = 0_{n}$\;
 \For{$i \gets 0$ \KwTo $n$}{
    $m[i] \gets i$\;
 }
 $p \gets 1$\;
 $\bTheta_1, m_1 \gets g(Q, \cA, m, p)$\;
 $\bTheta_2, m_2 \gets g(Q, \cA, m, p)$\;
 $\bz_1 \gets \textrm{mapback}(annealer(\bTheta_1,k), m_1)$\;
 $\bz_2 \gets \textrm{mapback}(annealer(\bTheta_2,k), m_2)$\;
$ f_1 \gets f_Q(Q,\bz_1); f_2 \gets f_Q(Q,\bz_2)$\;
  \eIf{$f_1<f_2$}{$\bz^* \gets \bz_1$; $f^* \gets f_1$; $m^* \gets m_1$ $\bz' \gets \bz_2$\;}{$\bz^* \gets \bz_2$ ; $f^* \gets f_2;$ $m^* \gets m_2$; $\bz' \gets \bz_1$\;}
\eIf{$f_1 \neq f_2$}{$S \gets \bz' \otimes \bz'-I_n+diag(\bz')$\;}{$S \gets 0_{n\times n}$\;}
$e \gets 0$; $d \gets 0$; $i \gets 0$; $\lambda\gets\lambda_0$\;
\Repeat{$i=i_{max}$ or ($e+d \geq N_{max}$ and $d<d_{min}$) }{
$Q' \gets Q+\lambda S$; \\
\If{$N$ divides $i$}{$p \gets p-\eta(p-p_\delta)$;}
$\bTheta', m \gets g(Q', \cA, m^*, p)$\;
$\bz' \gets \textrm{mapback}(annealer(\bTheta',k),m)$\;
with probability $q$: $\bz'$ $\rightarrow$ $h(\bz', p)$\;
  \eIf{$\bz' \not= \bz^*$} {
$f' \gets f_Q(Q,\bz')$;\\ 
 \eIf{$f'<f^*$}{$swap(\bz',\bz^*)$; $f^* \gets f'$; $m^* \gets m$; $e \gets 0$; $d \gets 0$\; 
 $S \gets S + \bz' \otimes \bz'-I_n+diag(\bz')$\;} 
 { $d \gets d+1$\;
  with probability $(p-p_\delta)^{(f'-f^*)}$: {
  $swap(\bz',\bz^*)$; $f^* \gets f'$; $m^* \gets m$; $e \gets 0$ \;}
 }
 $\lambda\gets \textrm{min}\Big\{\lambda_0,\frac{\lambda_0}{2+i-e}\Big\}$;
 }
 {$e \gets e+1$\;}
 $i \gets i+1$\;
 }
\Return $\bz^*$\;
\vspace{10pt}
\caption{Implementation of Quantum Annealing Tabu Search for QUBO problems.}
\label{algo:main}
\end{algorithm}
 
To test the correct functioning of the algorithm, a quantum annealer is required, but the available machine time at our disposal was very short. Hence, in order to avoid the usage of the quantum annealer while verifying the correctness of the implementations, the call to the annealer was replaced with an exhaustive search on the same equation that D-Wave minimizes: $E(\Theta, z) = \sum_{i\in V}\theta_i  z_i + \sum_{(i, j) \in E}\theta_{ij} z_i z_j$. By doing this, it was possible to understand if the development was going in the right way. Obviously, by minimizing the equation above, we were not simulating the behaviour of a quantum annealer since it is a probabilistic machine.

At the end of the correctness tests, we immediately noticed that the execution was very slow. The cause was the computation of the product of three matrices: the transpose $P^T$ of the permutation matrix $P$, the QUBO matrix $Q$ and the permutation matrix $P$. Indeed, the product $P^TQP$ \cite{QALS} is very expensive and inefficient in terms of computation; the time required to compute it could take up to 100 seconds. Since the permutation matrix $P$ is a permuted identity matrix and is very sparse, a sparse matrix representation was first chosen for it. In this way, it is possible to take into account only the non-zero values, storing triplets of $(row,\ column,\ value)$. The execution time for this computation was reduced to $\approx 1.5$ seconds, but there was still room for improvement.

To solve this efficiency problem, the representation of $P$ and the way of permuting the matrix $Q$ were changed. In detail, $P$ was replaced with a permutation vector $perm$, which represents the matrix $P$; let $P_i$ be the $i$th row in $P$. Here is an example:
\begin{equation*}
    P = \begin{bmatrix}
0 & 0 & 1 & 0 & 0\\ 
1 & 0 & 0 & 0 & 0\\ 
0 & 0 & 0 & 1 & 0\\ 
0 & 0 & 0 & 0 & 1\\ 
0 & 1 & 0 & 0 & 0
    \end{bmatrix}
    \qquad
    perm = [2, 0, 3, 4, 1]
\end{equation*}
\begin{center}
    \begin{alignat*}{1}
    &perm[0] = 2 \rightarrow P_0 \textrm{ has the value 1 in column 2}\\
    &perm[1] = 0 \rightarrow P_1 \textrm{ has the value 1 in column 0}\\
    &perm[2] = 3 \rightarrow P_2 \textrm{ has the value 1 in column 3}\\
    &perm[3] = 4 \rightarrow P_3 \textrm{ has the value 1 in column 4}\\
    &perm[4] = 1 \rightarrow P_4 \textrm{ has the value 1 in column 1}
    \end{alignat*}
\end{center}
The computational complexity of the permutation of $P$ is $O(n^2)$. Instead, the permutation of $perm$ has a complexity of $O(n\,log\,n)$, which is given by $O(n)$ accesses to the map $m$, each one with a complexity of $O(log\,n)$. By exploiting a hashmap, the time complexity could be reduced further to $O(n)$ on average.

After changing the representation of $P$, we investigated how to permute only the elements necessary for the mapping into the D-Wave topology. Let us consider an example matrix $Q$~\fnm{a}\fnt{a}{Note: $Q$ in this case is not a QUBO matrix, but just an example matrix.}, a permutation matrix $P$ and its equivalent permutation vector $perm$: 
\begin{equation*}
\label{eq:perm}
    Q = \begin{bmatrix}
1 & 2 & 3 & 4 & 5\\ 
6 & 7 & 8 & 9 & 10\\ 
11 & 12 & 13 & 14 & 15\\ 
16 & 17 & 18 & 19 & 20\\ 
21 & 22 & 23 & 24 & 25
\end{bmatrix}\qquad
    P = \begin{bmatrix}
0 & 0 & 0 & 1 & 0\\ 
1 & 0 & 0 & 0 & 0\\ 
0 & 0 & 0 & 0 & 1\\ 
0 & 1 & 0 & 0 & 0\\ 
0 & 0 & 1 & 0 & 0
\end{bmatrix}\qquad
    perm = [3, 0, 4, 1, 2]
\end{equation*}
The resulting matrix after the permutation ($P^TQP$) is the following:

\begin{equation*}
    M = \begin{bmatrix}
7 & 9 & 10 & 6 & 8\\
17 & 19 & 20 & 16 & 18\\
22 & 24 & 25 & 21 & 23\\
2 & 4 & 5 & 1 & 3\\
12 & 14 & 15 & 11 & 13
\end{bmatrix}
\end{equation*}
\noindent Then, let us see where each entry has moved after the permutation. The purpose is to find the element $m_{ij}$ without generating $M$. In practice, the indexes of the values $i$ and $j$ in the permutation vector $perm$ should be used to select the correct entry in the $Q$ matrix. For instance, to select the value for the entry $m_{1,2}$, the indexes of the values "1" and "2" in the permutation vector must be found:
\begin{itemize}
    \item "1" has index "3";
    \item "2" has index "4".
\end{itemize}
The first index identifies the row, while the second one identifies the column. It is now possible to find the entry $m_{1,2}$ by looking at the entry $q_{3,4}$, which is equal to 20.

Searching the index of a given value in the permutation vector has complexity $O(n)$; to generate the matrix $\Theta$, $O(n)$ searches are needed, because the D-Wave's Pegasus topology $A$ has $O(n)$ edges~\fnm{b}\fnt{b}{$n$ is the number of nodes of the Pegasus' sub-graph taken into consideration.}~\cite{pegasus}. Thus, the overall complexity is $O(n^2)$. However, since the objective is to find the index of several values, the best approach consists in inverting indexes and values. This operation can be done by running a $O(n)$ algorithm once at the beginning. The resulting complexity for the search operation becomes $O(1)$:
\begin{equation*}
    perm = [2, 0, 3, 4, 1]
    \qquad
    inverse = [1, 4, 0, 2, 3]
\end{equation*}
Let us look at the $inverse$ vector for the previous example: 

\vspace{-15pt}
\begin{alignat*}{1}
&inverse[0] = 1 \rightarrow perm[1] \textrm{ has the value 1 in column 0}\\
&inverse[1] = 4 \rightarrow perm[4] \textrm{ has the value 1 in column 1}\\
&inverse[2] = 0 \rightarrow perm[0] \textrm{ has the value 1 in column 2}\\
&inverse[3] = 2 \rightarrow perm[2] \textrm{ has the value 1 in column 3}\\
&inverse[4] = 3 \rightarrow perm[3] \textrm{ has the value 1 in column 4}
\end{alignat*}

Once the QALS algorithm \cite{QALS} has obtained an estimate of the solution from the quantum annealer, the variables in that solution must be mapped back to the original space of the problem (without permutation). Since the way of representing $P$ has changed, it is no longer possible to use $P^Tz$. Nevertheless, by exploiting the just described inverted permutation vector $inverse$, the original values can be obtained as $z\_back[i] \gets z[inverse[i]]$. The pseudocode is provided in \cref{alg:map-back}.

\begin{figure}[ht]
    \begin{algorithm}[H]
    \small
    \KwIn{Solution vector $z$, permutation vector $perm$}
    \KwResult{$perm^{-1}(z)$}
    $inverse \gets compute\_inverse(perm);$ \% computes the inverse of $perm$\\
    $z\_back \gets new\;int[n]$\;
    \For{$i \gets 0\;\textbf{to}\; n-1$}{
        $z\_back[i] \gets z[inverse[i]];$ \% $z\_back[i]$ takes the value contained in $z[inverse[i]]$, so that the values are mapped back to the original unpermuted space
    }
    \Return{$z\_back$}\;
    \vspace{10pt}
    \caption{Map back variables ($map\_back(z,\;perm)$ function).}
    \label{alg:map-back}
    \end{algorithm}
\end{figure}

\subsection{C++ implementation}
\label{subsec:c++-impl}
\vspace{5pt}

\paragraph{Lack of APIs}
At the time of writing, D-Wave provides only Python APIs. Therefore, a C++ program can not directly interface with D-Wave's QPUs. In particular, it is not possible to obtain the current QPU topology, submit problems and get the corresponding results. To solve these problems, two approaches have been taken into account.

The first approach consists in embedding a Python function in C++ \cite{embed}. This could be done by including the \texttt{Python.h} header and, after the creation of all required \texttt{PyObjects}, executing the Python function containing the calls to the quantum annealer. Nevertheless, the QALS algorithm is run on the Leap IDE, which does not provide the required \texttt{Python.h} header.

In the second approach (the one that has been used), at the startup, the C++ executable generates a Python child process by means of a fork operation. Its purposes is to provide the parent process with the QPU's topology and to submit each problem to the QPU's solver. In detail, the Python process first retrieves the annealer's topology and sends it to the C++ process via an anonymous pipe. This operation must be done since the quantum annealer could have inactive nodes, which must be taken into consideration during the execution (embedding). After receiving the topology, the parent process starts sending problems to the child one. In particular, for each entry in the matrix $\Theta$, the parent process sends the row index, the column index and the value; at the end, it sends a message containing a "\#" character to inform the receiver that all entries of the matrix have been sent. Meanwhile, the child process continuously reads the input pipe, storing the information in a dictionary. At the reception of the "\#" character, it submits the problem to the QPU, retrieves the solution and sends it back to the C++ process. Eventually, it starts reading the input pipe again.

\begin{algorithm}[b!]
    \small
    \KwIn{Number of measurements for each problem $k$ (\texttt{num\_reads})}
    $args \gets \{ "python",\;"solver.py",\;to\_string(k) \}$\;
    $dup2(fd[\texttt{READ}],\; \texttt{STDIN\_FILENO});$\% replace the standard input with an anonymous pipe\\
    $dup2(fd[\texttt{WRITE} + 2],\; \texttt{STDOUT\_FILENO});$\% replace the standard output with an anonymous pipe\\
    $close\_pipes(fd)$\;
    $execvp(args[0],\;args);$ \% replace the child's executable code
    \vspace{10pt}
    \caption{Initialization of the Python child process ($init\_child(k)$ function).}
    \label{alg:init}
\end{algorithm}

\begin{algorithm}[b!]
    \small
    \KwIn{Weights $\Theta$ ($\theta_{edge}$ refers to the weight associated to the $edge$ edge)}
    \KwResult{Vector with minimum estimated energy $z$}
    \% iterate over all edges, including pairs in the form $(i, i)$\;
    \ForEach{$edge\;\textbf{in}\;\Theta$}{
        $row \gets edge.u();$ \% store vertex $u$\\
        $col \gets edge.v();$ \% store vertex $v$\\
        $val \gets \theta_{edge};$ \% store weight associated to edge $(u,\;v)$\\
        $write(fd[\texttt{WRITE}],\;row);$ \% send row index\\
        $write(fd[\texttt{WRITE}],\;col);$ \% send column index\\
        $write(fd[\texttt{WRITE}],\;val);$ \% send value\\
    }
    $write(fd[\texttt{WRITE}],\;"\#")$ \% notify end of transmission\\
    \For{$i \gets 0\; \textbf{to}\; n-1$}{
        $z[i] \gets read(fd[\texttt{READ} + 2]);$ \\
    }
    \Return{$z$}\;
    \vspace{10pt}
    \caption{Send $\Theta$ to the Python process and retrieve the estimated solution $z$ \\ ($send\_to\_annealer(\Theta)$ function).}
    \label{alg:send}
\end{algorithm}

The pseudocode of two functions, i.e., \textit{init\_child} and \textit{send\_to\_annealer}, used to implement the last presented approach is shown in Algorithm \ref{alg:init} and \ref{alg:send}. In particular, the $init\_child$ function is in charge of creating the array of arguments to be passed to the Python process (using \texttt{dup2}), redirecting the standard input and output to pipes, and replacing the executable code with the Python one. \texttt{READ} (=0) and \texttt{WRITE} (=1) correspond to positions in the $fd$ array and identify which side of the pipe to use. Specifically, $fd[\texttt{READ}]$ is reserved for child reading, $fd[\texttt{WRITE}]$ for parent writing, $fd[\texttt{READ} + 2]$ for parent reading, and $fd[\texttt{WRITE} + 2]$ for child writing. Instead, $send\_to\_annealer$ sends each row - column - value triplet to the Python process and retrieves the solution from the pipe.

In order to determine how long the exchange of messages takes, we measured the time elapsed between the first message sent by the C++ process and the last message received by the same process, without actually calling the quantum annealer (the response is sent by the Python process). The transmission took $\approx$ 1 second using $5000$ variables, which is almost the Pegasus architecture limit. It is not too much, but it might cancel the benefit of using C ++ instead of Python.

\paragraph{Random numbers generation}
Looking at Algorithm~\ref{Alg}, it turns out that QALS \cite{QALS} relies on shuffling the map $m$ to find the best solution. To permute the map $m$, the following algorithm (Algorithm \ref{alg:shuffle}) has been used:

\begin{algorithm}[h!]
    \small
    \KwIn{Map $m$ ($m_x$ is the value of $m$ for key $x$)}
    \KwResult{Permuted map $m$}
    $shuffled \gets map()$\;
    $keys \gets new\;int [n]$\;
    \ForEach {$k \; \textbf{in} \; m.keys$}{
        $keys.append(k);$ \% create a vector of keys\\
    }
    $shuffle\_vector(keys);$ \% shuffle the vector of keys\\
    $it \gets keys.begin(\;);$ \% iterator\\
    \For {$pair \; \textbf{in} \; m$}{
        $shuffled_{pair.key} \gets m_{*it}$\;
        $it.next()$\;
    }
    \vspace{10pt}
    \caption{Shuffle map (\textit{shuffle(m)} function).}
    \label{alg:shuffle}
\end{algorithm}

\noindent The cornerstone of the algorithm is the shuffle of the map's keys. To properly work, the ideal shuffle algorithm for the $keys$ vector should be able to produce all the $n!$ permutations of $keys$. We decided to use the Fisher-Yates shuffle algorithm, which produces unbiased permutations, i.e., all permutations have the same probability.

The selected shuffle algorithm requires a random number generator; we initially decided to use C++ \texttt{rand()}. However, looking at the produced permutations, it turns out that the modulo operation (used to restrict the generated numbers to the required range) does not yield numbers with equal probability. Let us consider the example provided in \cite{rand}. The function \texttt{rand()} returns a number between 0 and \texttt{RAND\_MAX}. If we want to generate a random number between 0 and 2, $\texttt{rand()} \% 3$ will not necessarily produce each of the three values with equal probability. For instance, let us assume that $\texttt{RAND\_MAX} = 10$, then:
\begin{itemize}
    \item if \texttt{rand()} returns 0 or a multiple of 3, then $\texttt{rand()} \% 3 = 0$ and $P(0) = \frac{4}{11}$;
    \item if \texttt{rand()} returns 1, 4, 7 or 10, then $\texttt{rand()} \% 3 = 1$ and $P(1) = \frac{4}{11}$;
    \item if \texttt{rand()} returns 2, 5 or 8, then $\texttt{rand()} \% 3 = 2$ and $P(2) = \frac{3}{11}$.
\end{itemize}
In practice, the numbers between 0 and 2 do not have equal probability.

Another problem was the length of the period of the random number generator. Let us consider a deck of 52 cards; to create all the possible permutations, which are $52!$, a period that is at least equal to $52!$ is required. The period of \texttt{rand()} is typically $2^{32}$, but it depends on the implementations. Since the algorithm presented in \cite{QALS} has to permute vectors with more than $5000$ variables, \texttt{rand()} was not a reasonable choice.

A possible solution is to exploit the quantum annealer to generate random numbers. Since D-Wave's quantum annealer is a "trusted" quantum machine, if the weights are properly initialized, then the generated numbers are certainly random as they are produced by a quantum process. The idea is to submit one or more problems in which each qubit has the same probability to collapse to either 0 or 1. The quantum annealer will return an array of boolean values, which can be interpreted as one or more integer values (by splitting zero or more times the binary string representing the array). Here is the pseudocode for this approach:

\begin{algorithm}[h!]
    \small
    \KwIn{Amount of bits reserved per number $k$}
    \KwResult{Vector of random integers $nums$}
    $\Theta \gets \{ \; \};$ \% Python dictionary\\
    \For{$i \gets 0 \; \textbf{to} \; n-1$}{
        $\Theta[i][i] \gets 0;$ \% initialize the diagonal with all zeros\\
    }    
    $z \gets sample\_qubo(\Theta);$ \% run the annealer with $\Theta$\\
    $z \gets z.first.sample.values(\;);$ \% solution vector $z \in \{ 0, 1 \}^n$\\
    $nums \gets to\_decimal(z, k);$ \% extract $\lfloor n / k \rfloor$ decimal numbers, each one from 0 to $2^k - 1$\\
    \Return{$nums$}\;
    \vspace{10pt}
    \caption{Generation of random numbers by quantum annealing ($gen(k)$ function).}
    \label{alg:qrand}
\end{algorithm}

\noindent In detail, line 3 initializes the dictionary $\Theta$ with 0 values on the diagonal. Since all energy values are 0, any vector z minimizes Eq.~(\ref{costf}). Therefore, the quantum annealer will return any of the $2^n$ possible states with equal probability \cite{coins}. This method generates $\lfloor n/k \rfloor$ integer numbers at each call, where $n$ is the number of qubits used and $k$ is the amount of bits per number. For example, if $5000$ qubits are used to generate numbers from 0 to 127, one call will produce 714 integers (given by $\lfloor 5000/7 \rfloor$) in $\approx0.3$ seconds. It may seem good, but the cost is exaggerate:
\begin{itemize}
    \item at each iteration of QALS, a vector whose size can reach $\approx 5436$~\fnm{c}\fnt{c}{5436 is the number of available qubits at the time of writing.} must be permuted. Thus, at most $\approx 5436$ random numbers need to be generated at each iteration; this could be done using $n=5436$ and $k=13$. In a single run, Algorithm \ref{alg:qrand} would produce 418 numbers with the considered parameters; therefore, to generate $5436$ integers, it should be executed 13 times;
    \item each run of Algorithm \ref{alg:qrand} requires around 0.3 seconds (on average). As a consequence, the overhead for each iteration of QALS would be around 3.9 seconds.
\end{itemize}
In practice, the quantum annealer would be called 13 times more than usual, reducing the amount of possible experiments or, in other perspective, increasing the cost. Obviously, it is not a feasible approach.

In the end, we chose the Mersenne Twister, which uses a period of $2^{19937} - 1$. This does not guarantee quality in random numbers generation but certainly allows for longer sequences than \texttt{rand()}'s period. In \cite{mt}, the authors claim that the Mersenne Twister creates 64-bit floating point random numbers faster than the hardware-implemented Intel Secure Key. Although it is not sufficient to produce all possible sequences, it is the best compromise we found between efficiency and coverage of the space of permutations.

\subsection{Python implementation}
\label{subsec:python-impl}
\vspace{5pt}

\begin{algorithm}[b!]
\small
\KwIn{Sampler $sampler$, QUBO problem $Q$, dimension of QUBO problem $n$}
\KwResult{Embedded QUBO problem $mapped$} 
$actives \gets dict()$\;
\ForEach{$node \in sampler.nodelist$}{
    $actives[node] \gets list()$\;
    \If{$actives.keys().size() == n$}{
        \textbf{break};
    }
}
\ForEach{$edge \in sampler.edges$}{ 
    \If{$edge.node1 \in actives.keys()$ \textbf{and} $\  edge.node2 \in actives.keys()$}{
        $actives[edge.node1].append(edge.node2)$\;
        $actives[edge.node2].append(edge.node1)$\;
    }
}
$support_n \leftarrow dict()$\;
$i \leftarrow 0$\;
\ForEach{$node \in actives.keys().ordered()$}{
    $support[node] = i$\;
    $i \leftarrow i + 1$\;
}
$mapped_{n\times n} \leftarrow 0$\;
\ForEach{$node \in actives.keys()$}{
    \ForEach{$adjnode \in actives[node]$}{
        $mapped$[$node$][$adjnode$] = $Q$[$support$[$node$]][$support$[$adjnode$]]\;
    }
}
\Return $mapped$\;
\vspace{10pt}
\caption{Embedding the problem in the sampler topology.}
\label{algo:embedding}
\end{algorithm}

\paragraph{Embedding}
Embedding the QUBO matrix in the Pegasus topology is, at least in theory, straightforward: we have just to assign a $node \in V$ to each index of the QUBO matrix, where $V$ is the list of nodes in the topology, and map the entries accordingly. However, the graph is not complete and some qubits are not available (97$\%$ of the total number of qubits is actually available). D-Wave, aware of the problem, provides a function that lists all the working qubits and edges; for instance, the function returns only 5436 qubits for the Pegasus topology. In practice, it is possible to embed the QUBO matrix in the topology using a custom algorithm or the \texttt{EmbeddingComposite} class, which is part of the D-Wave's APIs. However, \texttt{EmbeddingComposite} has an intrinsic limit: it works for QUBO matrices of size up to 196 $\times$ 196. Therefore, a custom algorithm was the best choice for us. The structure of the algorithm we have used, whose pseudocode is shown in Algorithm \ref{algo:embedding}, is the following:
\begin{enumerate}
    \item it initializes an empty dictionary using the required number of working nodes as keys;
    \item it associates each key (working node) with the list of nodes that are the endpoint of a working edge outgoing from the considered node (the nodes in the lists must be keys of the same dictionary);
    \item it creates a support dictionary that maps each $node$ to an index of the QUBO matrix;
    \item it embeds the QUBO matrix by looping on each $node$ (that corresponds to a row of the matrix) and its $adjacency\ nodes$ (that correspond to the columns of the matrix).
\end{enumerate}
It is worth mentioning that this custom algorithm has been implemented in Python and used also in the C++ implementation.

\paragraph{Communication with the annealer}
D-Wave provides several samplers to solve problems. In particular, the samplers share the method signatures; the difference lies in the internal implementations. As regards the algorithm used to submit problems to the annealer (Algorithm \ref{algo:script}), it is very simple, but it is necessary to highlight some aspects: 
\begin{enumerate}
    \item $\Theta$ must be provided as a dictionary, otherwise the sampler will not process the submitted QUBO problem; 
    \item $k$ represents the number of annealer reads (as in Algorithm \ref{algo:main}). It is optional for simulated and quantum annealing, whereas Hybrid does not support it.
\end{enumerate}

\begin{algorithm}[t!]
\small
\KwIn{Dictionary representing the embedding of the problem in the topology $\Theta$, sampler $sampler$, number of reads $k$}
\KwResult{List of $\{0,1\}^n$}
$response = sampler.sample\_qubo(\Theta, num\_reads=k)$\;
\Return $response.first.sample.values()$\;
\vspace{10pt}
\caption{Communication with the annealer (problem submission).}
\label{algo:script}
\end{algorithm}

\section{Experimental Results}
\label{sec:experimental-results}
\noindent
Two problems have been selected to test the performance of QALS:
\begin{itemize}
    \item the Number Partitioning Problem (NPP);
    \item the Travelling Salesman Problem (TSP).
\end{itemize}
In all NPP tests, the results for the approaches that make use of the quantum annealer have been obtained with a single (or rarely double) run. Indeed, the time available on the quantum annealer was too little to perform more of them. Instead, in TSP, we have performed multiple runs due to the less QPU usage per run.

\paragraph{QALS parameters legend}
\vspace{-8pt}
\begin{itemize}
    \item $p_{\delta}$ is the minimum probability of permutation modification;
    \item $\eta$ is the probability decreasing rate;
    \item $q$ is the candidate perturbation probability;
    \item $N$ is the number of iterations at constant probability;
    \item $\lambda_0$ is the initial balancing factor for the tabu matrix;
    \item $k$ is the number of annealer measurements for each problem submitted to it;
    \item $N_{max}$ is the maximum number of consecutive times that QALS can find the previous solution or a solution that is not better than $f^*$ before stopping;
    \item $d_{min}$ represents a further condition on the number of times that QALS can find a solution worse than $f^*$ (it must be less than $d_{min}$) before stopping.
\end{itemize}

\subsection{Number Partitioning Problem (NPP)}
\label{subsubsec:npp}

\paragraph{Definition}
The Number Partitioning Problem (NPP) consists in splitting a set of numbers into two subsets such that the difference between the sum of the values in the first subset and the sum of the values in the second subset is minimum. The QUBO formulation of NPP presented here and used in the experiments has been taken from \cite{glover2019tutorial}. In detail, consider a set of numbers $S$ = \{$s_1$, $s_2$, ..., $s_n$\}. If $s_i$ is assigned to the first subset, $x_i$ = 1; otherwise, $x_i$ = 0. Then, the sum of the values in the first subset is given by $sum_1$ = $\sum_{i=1}^{n} s_i x_i$, whereas the sum of the values in the second subset is equal to $sum_2$ = $\sum_{i=1}^{n} s_i - \sum_{i=1}^{n} s_i x_i$. Let $c$ be equal to $\sum_{i=1}^{n} s_i$, thus the difference between the two sums is:
\begin{gather}
    \textrm{\textit{diff}} = \displaystyle\sum_{i=1}^{n} s_i - 2\sum_{i=1}^{n} s_i x_i = c - 2\sum_{i=1}^{n} s_i x_i
\end{gather}
Instead of directly minimizing the difference, let us minimize its value squared, which can be written as follows: 
\begin{gather}
    \textrm{\textit{diff}}^2 = \Bigg\{c-2\displaystyle\sum_{i=1}^{n}s_i x_i\Bigg\}^2 = c^2 + 4x^tQx
\end{gather}
Here, $Q$ is the QUBO matrix, whose entries are given by:
\begin{gather}
    Q_{ii} = s_i\left(s_i - c\right) \hspace{10pt} Q_{ij} = Q_{ji} = s_i s_j
\end{gather}

\noindent By dropping the constants ($c^2$ and $4$), the QUBO optimization problem can be defined as:
\begin{gather}
    QUBO: \textrm{min } y = x^tQx
\end{gather}

\noindent Algorithm \ref{alg:npp} provides the pseuodocode for the computation of the entries of the QUBO matrix.

\begin{algorithm}[b!]
    \small
    \KwIn{Set of numbers $s$}
    \KwResult{Matrix $Q$ representing the QUBO formulation of the NPP problem defined by $s$}
    $n \gets s.size()$\;
    $c \gets sum(s);$ \% sum over numbers in $s$\\
    $Q \gets new\;int[n][n]$\;
    \For{$i \gets 0$ \textbf{to} $n-1$}{
        \For{$j \gets 0$ \textbf{to} $n-1$}{
            \eIf{$i \neq j$}{
                $p \gets s[i] \cdot s[j]$\;
                $Q[i][j] \gets p$\;
                $Q[j][i] \gets p$\;
            }{
                $Q[i][i] \gets s[i] \cdot (s[i] - c)$\;
            }
        }
    }
    \Return{$Q$}\;
    \vspace{10pt}
    \caption{Translation of NPP into QUBO.}
    \label{alg:npp}
\end{algorithm}

\paragraph{Example} Consider the following set of eight numbers:
\begin{equation*}
    S = [8, 21, 6, 7, 16, 9, 10, 27]
\end{equation*}
It follows that $c^2$ = 10816 and the equivalent QUBO problem is min $y = x^tQx$ with:
\vspace{10pt}
\begin{center}$Q$ =
    $\begin{bmatrix}
    -768 &  168 &  48 &  56 &  128 &  72 &  80 &  216\\
    168 &  -1743 &  126 &  147 &  336 &  189 &  210 &  567\\
    48 &  126 &  -588 &  42 &  96 &  54 &  60 &  162\\
    56 &  147 &  42 &  -679 &  112 &  63 &  70 &  189\\
    128 &  336 &  96 &  112 &  -1408 &  144 &  160 &  432\\
    72 &  189 &  54 &  63 &  144 &  -855 &  90 &  243\\
    80 &  210 &  60 &  70 &  160 &  90 &  -940 &  270\\
    216 &  567 &  162 &  189 &  432 &  243 &  270 &  -2079
    \end{bmatrix}$
\end{center}
\vspace{10pt}
The solution to the QUBO problem is $x$ = (1,1,1,1,0,0,1,0), for which $y$ = -2704 and:
\begin{equation*}
    \textrm{\textit{diff}}^2 = c^2 + 4x^tQx = 10816 + 4 \cdot (-2704) = 10816 - 10816 = 0
\end{equation*}

\paragraph{Classical algorithms for NPP}
Since NPP is a NP-hard problem, no efficient algorithm to solve it has been found yet. Instead, several heuristics have been proposed, such as the greedy heuristic, which consists in sorting the numbers in descending order and adding each of them to the set whose value is the smaller so far, or the Karmarkar-Karp (KK) one \cite{kk}. The latter is the basis of an exact (and exponential time) algorithm originally proposed in \cite{Korf98} and tested in \cite{Pedroso2013}, i.e., the Complete Karmarkar-Karp algorithm (CKK). 

In detail, CKK performs a depth-first search of a binary tree in which the left branch corresponds to replacing the two largest numbers at the current level with the absolute value of their difference (i.e., the numbers are placed in different subsets), whereas the right branch corresponds to replacing them with their sum (i.e., they are placed in the same subset). However, the subset in which each number goes is not fixed during the search. Indeed, the resulting subsets are retrieved only at the end of the search by running a linear time procedure. The complexity of the algorithm is exponential in the worst case; nevertheless, the search stops if a perfect partition is found, i.e., the last remaining number, which represents the difference between the two subsets, is equal to either 0 or 1. Moreover, if the current largest number is greater than the sum of all the others, the branching for that sub-tree can be stopped (in this case, all the other numbers can be placed in the same subset). Eventually, with four numbers or less, only the left branch should be taken into account since the KK heuristic, which corresponds to the left branch action, is exact in this case.

In this work, the implementation of CKK provided by \cite{Pedroso2013} has been used for comparison in the experiments.  

\paragraph{Experimental procedure}
The runs have been launched on the Leap cloud from the command line. Moreover, they have been executed one at a time due to some issues in terms of computational time, which grows with multiple parallel computations. 

\paragraph{Results}
The following values have been used for QALS parameters in all runs: 

\begin{table}[t!]
    \tcaption{Values used for QALS parameters in all NPP tests.}
    \label{tab:data_tests_npp}
    \centerline{\footnotesize
    \begin{tabular}{|c|c|c|c|c|c|c|c|c|c|}
        \hline
        $p_\delta$ & $\eta$ & $q$ & $N$ & $\lambda_0$ & $k$ & $N_{max}$ & $d_{min}$ \\
        \hline
        0.1 & 0.01 & 0.2 & 10 & 1.5 & 10 & 100 & 70 \\
        \hline
    \end{tabular}}
\end{table}

\noindent Concerning the annealing parameters, such as the annealing time and schedule, the default values have been used for QALS (the system used is Advantage 1.1) \cite{solver_params}; instead, Hybrid autonomously manages these properties. Actually, all NPP results have been obtained using the C++ implementation, whereas the Python implementation has been exploited to check them. The results are shown in \cref{tab:npp-res}. In particular, \textit{Range} represents the upper limit of the number generation interval (~$[1,\ Range]$~), thus, the maximum possible $s_i$ value in the experiment; for instance, if \textit{Dimension} is 500 and \textit{Range} is 100, a vector of 500 integer numbers in the range $[1, 100]$ is used. Instead, \textit{Sets Difference} corresponds to the difference between the two resulting sets.

\paragraph{NPP results discussion}
The QALS algorithm did not perform as well as expected. Indeed, Hybrid turned out to be better in terms of results and faster in terms of computational time. Moreover, the classical algorithm used for comparison proved to be very fast and always returned the best solution for the given set \textit{S}. In general, classical algorithms for the Number Partitioning Problem are efficient even if the problem is NP-Hard. The performance of QALS could probably improve by considering a much higher number of annealer measurements ($k \gg 10$). Indeed, the D-Wave quantum annealer is subjected to noise and temperature effects, and larger output samples would increase the probability of obtaining a solution close to the optimum. Nevertheless, the time at our disposal was too little to consider higher $k$ values.

\begin{table}[b!]
\tcaption{Tests performed on the Number Partitioning Problem.}
\label{tab:npp-res}
\centerline{\footnotesize
\begin{tabular}{|c|c|c|c|c|c|}
        \hline
        \textbf{Dimension} & \textbf{Approach} & \textbf{Range} & \textbf{Sets Difference} & \textbf{Time}(s) & \textbf{\# Iterations} \\
        \hline
        \multirow{12}{*}{500} & QALS & \multirow{3}{*}{100} & 93 & 2172 & 4000\\
        & Hybrid & & 1 & 6 & -\\
        & Classical & & 1 & 0.003 & -\\
        \cline{2-6}
        & QALS & \multirow{3}{*}{1000} & 292 & 2157 & 4000 \\
        & Hybrid & & 4 & 9 & -\\
        & Classical & & 0 & 0.005 & -\\
        \cline{2-6}
        & QALS & \multirow{3}{*}{10000} & 2640 & 2151 & 4000 \\
        & Hybrid & & 36 & 12 & -\\
        & Classical & & 0 & 0.004 & -\\
        \cline{2-6}
        & QALS & \multirow{3}{*}{1000000} & 475860 & 1992 & 4000 \\
        & Hybrid & & 2340912 & 14 & -\\
        & Classical & & 0 & 0.005 & -\\
        \hline
        \hline
        \multirow{12}{*}{1200} & QALS & \multirow{3}{*}{1000} & 185 & 1327 & 2000\\
        & Hybrid & & 1 & 23 & -\\
        & Classical & & 1 & 0.015 & -\\
        \cline{2-6}
        & QALS & \multirow{3}{*}{10000} & 6337 & 1304 & 2000\\
        & Hybrid & & 225 & 23 & -\\
        & Classical & & 1 & 0.017 & -\\
        \cline{2-6}
        & QALS & \multirow{3}{*}{100000} & 145982 & 1270 & 2000\\
        & Hybrid & & 186624 & 23 & -\\
        & Classical & & 0 & 0.024 & -\\
        \cline{2-6}
        & QALS & \multirow{3}{*}{1000000} & 303833 & 1267 & 2000\\
        & Hybrid & & 781440 & 23 & -\\
        & Classical & & 0 & 0.019 & -\\
        \hline
\end{tabular}}
\end{table}

\begin{table}[t!]
\caption*{\cref{tab:npp-res} (Continued): Tests performed on the Number Partitioning Problem.}
\centerline{\footnotesize
\begin{tabular}{|c|c|c|c|c|c|}
        \hline
        \textbf{Dimension} & \textbf{Approach} & \textbf{Range} & \textbf{Sets Difference} & \textbf{Time}(s) & \textbf{\# Iterations} \\
        \hline
        \multirow{12}{*}{2500} & QALS & \multirow{3}{*}{1000} & 3108 & 2442 & 2000\\
        & Hybrid & & 0 & 62 & -\\
        & Classical & & 0 & 0.072 & \\
        \cline{2-6}
        & QALS & \multirow{3}{*}{10000} & 11681 & 2666 & 2000\\
        & Hybrid & & 25 & 54 & -\\
        & Classical & & 1 & 0.090 & -\\
        \cline{2-6}
        & QALS & \multirow{3}{*}{100000} & 160676 & 2354 & 2000\\
        & Hybrid & & 6240 & 64 & -\\
        & Classical & & 1 & 0.089 & -\\
        \cline{2-6}
        & QALS & \multirow{3}{*}{1000000} & 2731518 & 2341 & 2000\\
        & Hybrid & & 1151232 & 65 & -\\
        & Classical & & 1 & 0.093 & -\\
        \hline
        \hline
        \multirow{12}{*}{5436} & QALS & \multirow{3}{*}{1000} & 6209 & 9414 & 2000\\
        & Hybrid & & 1 & 260 & -\\
        & Classical & & 1 & 0.318 & -\\
        \cline{2-6}
        & QALS & \multirow{3}{*}{10000} & 528 & 9413 & 2000\\
        & Hybrid & & 16 & 267 & -\\
        & Classical & & 0 & 0.437 & -\\
        \cline{2-6}
        & QALS & \multirow{3}{*}{100000} & 4010004 & 9168 & 2000\\
        & Hybrid & & 12112 & 263 & -\\
        & Classical & & 0 & 0.464 & -\\
        \cline{2-6}
        & QALS & \multirow{3}{*}{1000000} & 5497085 & 9507 & 2000\\
        & Hybrid & & 24576 & 260 & -\\
        & Classical & & 0 & 0.479 & -\\
        \hline
\end{tabular}}
\end{table}
    
\subsection{Travelling Salesman Problem (TSP)}
\label{subsubsec:tsp}

\paragraph{Definition}
The Travelling Salesman Problem, also known as TSP, consists in finding the shortest route through $n$ cities, given the list of cities and their pairwise distances (each city is not necessarily directly connected to all the others). In detail, the route must visit each city exactly once and must return to the origin city. Hence, TSP can be seen also as the problem of finding, for a graph $G$ = ($V$, $E$), the Hamiltonian cycle such that the sum of the weights of the edges in the cycle is minimum. This last perspective is particularly relevant since, as presented in \cite{Lucas_2014}, there exists a QUBO formulation for it. Specifically, the problem Hamiltonian for TSP is the following:
\begin{gather}
    \nonumber H = H_A + H_B \\
    \nonumber H_A = A\sum_{i=1}^{n}\left(1-\sum_{j=1}^{n}x_{i,j}\right)^2 + A\sum_{j=1}^{n}\left(1-\sum_{i=1}^{n}x_{i,j}\right)^2 +
    A\sum_{(uv)\notin E}\sum_{j=1}^{n}x_{u,j}x_{v,j+1} \\
    H_B = B\sum_{(uv)\in E}W_{uv}\sum_{j=1}^{n}x_{u,j}x_{v,j+1}
\end{gather}
where $x_{i,j}$ is 1 if the node (city) $i$ is in position $j$ in the cycle (route), 0 otherwise, $x_{v,n+1} = x_{v,1}$, and $A$ and $B$ are positive constants ($A,B > 0$). Here, $H_A$ encodes the constraints of the problem, i.e., each node appears exactly once in the cycle (first term), there is only one node in each position of the cycle (second term), and the order of nodes in the cycle is valid (third term). Instead, $H_B$ encodes the minimization of the cycle total weight (route length); indeed, $W_{u,v}$ is the weight of the $(u,v)$ edge. Finally, in order to make not favourable to violate the constraints, the following relationship must be satisfied: $0 < B(max(W_{uv})) < A$. In the experiments, $B$ was set to $1$ and $A$ was set to $n \times  max(W_{uv})$.

\clearpage

\begin{algorithm}[t!]
    \small
    \KwIn{Distance matrix $D$}
    \KwResult{Matrix $Q$ representing the QUBO formulation of the TSP problem defined by $D$}
    $n \gets size(D, 0);$ \% number of rows of D (square matrix) \\
    $Q \gets new\;int[n^2][n^2]$\;
    $all\_zeros(Q)$\;
    $A \gets n \cdot max\_coeff(D);$ \% penalty set according to \cite{Lucas_2014} \\
    $B \gets 1;$ \% multiplier set according to \cite{Lucas_2014} \\
    $Q \gets add\_cost\_objective(Q, D, B)$\;
    $Q \gets add\_time\_constraints(Q, A)$\;
    $Q \gets add\_position\_constraints(Q, A)$\;
    \Return{$Q$}\;
    \vspace{10pt}
    \caption{Translation of TSP into QUBO \cite{tsp_implementation}.}
    \label{alg:tsp}
\end{algorithm}

\begin{algorithm}[t!]
    \small
    \KwIn{QUBO matrix $Q$, distance matrix $D$, multiplier $B$}
    \KwResult{Matrix $Q$ including the cost objective}
    $n \gets size(D, 0);$ \% number of rows of D (square matrix) \\
    \For{$t \gets 0$ \textbf{to} $n-1$}{
        \For{$i \gets 0$ \textbf{to} $n-1$}{
            \For{$j \gets 0$ \textbf{to} $n-1$}{
                $r \gets t \cdot n + i$\;
                $c \gets (t + 1)$ mod $n^2 + j$\;
                $Q[r][c] \gets B \cdot D[i][j]$\;
            }
        }
    }
    \Return{$Q$}\;
    \vspace{10pt}
    \caption{$add\_cost\_objective$ function (computation of $H_B$ coefficients).}
    \label{alg:tsp.cost}
\end{algorithm}

\begin{algorithm}[t!]
    \small
    \KwIn{QUBO matrix $Q$, constraint penalty $A$}
    \KwResult{Matrix $Q$ including time constraints}
    $n \gets size(D, 0);$ \% number of rows of D (square matrix) \\
    \For{$t \gets 0$ \textbf{to} $n-1$}{
        \For{$i \gets 0$ \textbf{to} $n-1$}{
            $r \gets t \cdot n + i$\;
            $Q[r][r] \gets Q[r][r] - A$\;
            \For{$j \gets 0$ \textbf{to} $n-1$}{
                \If{$i \neq j$}{
                    $c \gets t \cdot n + j$\;
                    $Q[r][c] \gets 2 \cdot A$\;
                }
            }
        }
    }
    \Return{$Q$}\;
    \vspace{10pt}
    \caption{$add\_time\_constraints$ function (computation of part of $H_A$ coefficients).}
    \label{alg:tsp.time}
\end{algorithm}

\clearpage

Given the problem Hamiltonian $H$, the variables $x_{i,j}$ must be renumbered in order to build the QUBO matrix $Q$. In detail, the following renumbering was applied here:
\begin{equation}
    (x_{1,1},\ x_{1,2},\ ...\ x_{1,n},\ x_{2,1}\ ...\ x_{n,n}) \rightarrow (x_{1},\ ...\ x_{n^2})
\end{equation}
At this point, the entry $Q_{ij}$ of $Q$ for $i,j \in \{1, ... n^2\}$ is just the coefficient of $x_i x_j$ in $H$.

The pseudocode for the computation of the entries of the QUBO matrix is provided in Algorithms \ref{alg:tsp}-\ref{alg:tsp.cost}-\ref{alg:tsp.time}-\ref{alg:tsp.position}.

\paragraph{Experimental procedure}
As for the Number Partitioning Problem, all runs have been launched on the Leap cloud from the command line and have been executed one at a time. The main reason is the same, but there is another one here: Hybrid would fail if launched in parallel with other Hybrid processes.

\begin{algorithm}[t!]
    \small
    \KwIn{QUBO Matrix $Q$, constraint penalty $A$}
    \KwResult{Matrix $Q$ including position constraints}
    $n \gets size(D, 0);$ \% number of rows of D (square matrix) \\
    \For{$i \gets 0$ \textbf{to} $n-1$}{
        \For{$t1 \gets 0$ \textbf{to} $n-1$}{
            $r \gets t1 \cdot n + i$\;
            $Q[r][r] \gets Q[r][r] - A$\;
            \For{$t2 \gets 0$ \textbf{to} $n-1$}{
                \If{$t1 \neq t2$}{
                    $c \gets t2 \cdot n + i$\;
                    $Q[r][c] \gets 2 \cdot A$\;
                }
            }
        }
    }
    \Return{$Q$}\;
    \vspace{10pt}
    \caption{$add\_position\_constraints$ function (computation of part of $H_A$ coefficients).}
    \label{alg:tsp.position}
\end{algorithm}

\paragraph{Results}
The following values have been used for QALS parameters in all runs:
\begin{table}[ht]
    \tcaption{Values used for QALS parameters in all TSP tests.}
    \label{tab:data_tests_tsp}
    \centerline{\footnotesize
    \begin{tabular}{|c|c|c|c|c|c|c|c|c|c|}
        \hline
        $p_\delta$ & $\eta$ & $q$ & $N$ & $\lambda_0$ & $k$ & $N_{max}$ & $d_{min}$ \\
        \hline
        0.1 & 0.2 & 0.2 & 5 & 1.5 & 5 & 100 & 70 \\
        \hline
    \end{tabular}}
\end{table}

\noindent In order to make a fair comparison, the same number of annealer measurements ($k=5$) has been used for Embedding Composite. Regarding the annealing parameters (like the annealing time and schedule), the default values have been used for both Embedding Composite and QALS (the system used is Advantage 1.1) \cite{solver_params}; as mentioned previously, Hybrid autonomously manages these properties. The results for Hybrid, Embedding Composite and QALS have been obtained mainly using the Python implementation, exploiting the C++ implementation to check them; instead, Brute Force has been executed using a C++ implementation. All results are reported in Table \ref{tab:tsp-res}. In detail, \pmb{$\mu$} is the average TSP cost across runs, \pmb{$\si$} is the corresponding standard deviation, and the average time is reported in seconds. The TSP cost has been computed as the length of the route (sum of the weights of the edges in the cycle) in the original problem space. Eventually, it is worth highlighting that, in all these experiments, the $W_{u,v}$ edge weights (city distances) are real numbers in the range $[0, 10]$.

\begin{table}[t!]
    \tcaption{Tests performed on the Travelling Salesman Problem.}
    \label{tab:tsp-res}
    \centerline{\footnotesize
    \begin{tabular}{|c|c|c|c|c|c|c|}
        \hline
        \textbf{TSP Size} & \textbf{QUBO Size} & \textbf{Approach} & \pmb{$\mu$} & \pmb{$\si$} & \textbf{Avg. Time (s)} & \textbf{\# Runs} \\
        \hline
        \multirow{4}{*}{10} & \multirow{4}{*}{100} & Brute Force & 35.20 & 0 & 48 & 10\\
        \cline{3-7}
        & & Hybrid & 36.94 & 1.16 & 13.78 & 3 \\
        \cline{3-7}
        & & E.C. with S.R. & 55.13 & 3.22 & 94.66 & 10 \\
        \cline{3-7}
        & & QALS with S.R. & 49.95 & 3.02 & 64.83 & 10 \\
        \hline
        \hline
        \multirow{4}{*}{12} & \multirow{4}{*}{144} & Brute Force & 26.21 & 0 & 56.8 & 3\\
        \cline{3-7}
        & & Hybrid & 33.43 & 1.26 & 14.86 & 10 \\
        \cline{3-7}
        & & E.C. with S.R. & 52.91 & 6.5 & 132.96 & 10\\
        \cline{3-7}
        & & QALS with S.R. & 54.77 & 4.53 & 265.3 & 10\\
        \hline
        \hline
        \multirow{4}{*}{14} & \multirow{4}{*}{196} & Brute Force & 33.94 & 0 & 10630 & 3\\
        \cline{3-7}
        & & Hybrid & 49.48 & 1.99 & 15.7 & 3 \\
        \cline{3-7}
        & & E.C. with S.R. & 67.50 & 9.87 & 465 & 3 \\
        \cline{3-7}
        & & QALS with S.R. & 74.58 & 7.87 & 180 & 10 \\
        \hline
        \hline
        \multirow{4}{*}{32} & \multirow{4}{*}{1024} & Brute Force & - & - & - & - \\
        \cline{3-7}
        & & Hybrid & 124.72 & 3.45 & 24.98 & 3 \\
        \cline{3-7}
        & & E.C. with S.R. & - & - & - & - \\
        \cline{3-7}
        & & QALS with S.R. & 157.99 & 10.23 & 588 & 10 \\
        \hline
        \hline
        \multirow{4}{*}{64} & \multirow{4}{*}{4096} & Brute Force & - & - & - & - \\
        \cline{3-7}
        & & Hybrid & 288.87 & 5.82 & 23.25 & 3 \\
        \cline{3-7}
        & & E.C. with S.R. & - & - & - & -\\
        \cline{3-7}
        & & QALS with S.R. & 336.94 & 17.9 & 3570 & 10 \\
        \hline
        \hline
        \multirow{4}{*}{72} & \multirow{4}{*}{5184} & Brute Force & - & - & - & - \\
        \cline{3-7}
        & & Hybrid & 331.24 & 16.39 & 37.36 & 3 \\
        \cline{3-7}
        & & E.C. with S.R. & - & - & - & - \\
        \cline{3-7}
        & & QALS with S.R. & 387.00 & 21.25 & 2528 & 3 \\
        \hline
        \hline
        \multirow{4}{*}{74} & \multirow{4}{*}{5476} & Brute Force & - & - & - & - \\
        \cline{3-7}
        & & Hybrid & 344.19 & 2.8 & 38.65 & 3 \\
        \cline{3-7}
        & & E.C. with S.R. & - & - & - & - \\
        \cline{3-7}
        & & QALS with S.R. & - & - & - & - \\
        \hline
    \end{tabular}}
    \vspace{15pt}
    \centerline{\footnotesize E.C. = Embedding Composite \quad \quad S.R. = Solution Refinement}
\end{table}

\paragraph{Solution refinement}
The solutions found by QALS and Embedding Composite without refinement have never satisfied the $H_A$ constraints, as opposed to the ones generated by Hybrid. In particular, Embedding Composite has no guarantees to find the best solution or to find a solution that satisfies the given constraints. Regarding QALS, since only a subpart of the entire problem (including the encoding of the constraints) is mapped to the annealer at each iteration, the constraints are usually not fully encoded in the topology and, as a consequence, turn out to be satisfied only if the optimal solution is found or in a few other cases~\fnm{d}\fnt{d}{These statements are based on the empirical results obtained.}. In practice, both Embedding Composite and QALS have never produced valid TSP solutions, hence their results without refinement have not been reported here. 

A TSP solution in QUBO formulation is represented by an integer sequence of length $n^2$, where $n$ is the number of nodes (cities); the solution can be also seen as $n$ sub-sequences of $n$ locations each. In order to be a valid solution, each sub-sequence must contain $n-1$ 0s and exactly one 1; moreover, the position of the value 1 inside the $i$-th sub-sequence must be unique w.r.t. all sub-sequences. For instance, this is a valid solution for three cities:
\begin{equation*}
    x_1 = [0, 1, 0,\;1, 0, 0,\; 0, 0, 1] 
\end{equation*}
whereas this is not:
\begin{equation*}
    x_2 = [1, 1, 0,\;1, 0, 0,\; 0, 0, 0] 
\end{equation*}
In order to reduce a QUBO TSP solution to a classical TSP one, each sub-sequence is seen as an integer number corresponding to the position of the value 1. For example, the classical version of $x_1$ is:
\begin{equation*}
    s_1 = [1, 0, 2] 
\end{equation*}
Instead, it is not possible to reduce $x_2$ to a valid classical TSP solution unless it is refined, i.e., modified a bit.

To turn a non valid QUBO TSP solution into a valid classical TSP one while maintaining as intact as possible the original solution, the following steps have been applied. Let $x$ be the annealer's solution vector, $f: {\mathbb B}^n\rightarrow {\mathbb N}^k$ be a function that returns a vector of size $k$ containing the position of all values different from 0 in $x_i$ (with $x_i$ being the $i$-th sub-vector of length $n$ in $x$), $A$ be the set $\{0, ..., n-1\}$. First, a solution vector of length $n$ is initialized with -1 values, i.e., $s \in \{-1\}^n$. Let $R$ be a vector of sets, where each set $R_i = \{f(x_i)\}$ contains the possible solutions (nodes) for the $i$-th sub-sequence (based on the annealer's solution vector); then, $s_i = R_i[0] \iff sizeof(R_i) = 1$. Let us define the set of unavailable nodes $K$.

\begin{definition}
An unavailable node is a node that represents the solution for a certain cycle position (i.e., there is no other 1 value in the corresponding sub-sequence) or a node marked as such after randomly picking it from the available ones. In the following example, node 0 is unavailable.
\begin{equation*}
    x_2 = [1, 1, 0,\;1, 0, 0,\; 0, 0, 0] 
\end{equation*}
\end{definition}

\noindent In practice, $K = \{s_i \;|\; s_i \neq -1, \;i \in  \{0, ...,n-1\}\}$. At this point, let $D_i$ be $R_i \setminus K$. The next step consists in computing $D_i$ for the first $i$ in $\{0,..., n-1\}$ such that $sizeof(R_i) > 1$. If $sizeof(D_i) > 0$, a random value $d \in D_i$ is chosen and the following operations are performed: $s_i = d$ and $K = \{K \cup d\}$, i.e., the chosen available value is assigned to the considered position of the solution vector and the set of unavailable nodes is updated accordingly. The procedure is then repeated for all subsequent $i$ values satisfying the condition on the size of $R_i$.\\ 
After these operations, a vector $idx_i$ is created for each node $i$ in $\{0,..., n-1\}$. In particular, $j \in idx_i \iff s_j = i$ (here, $j \in \{0,..., n-1\}$); hence, $idx_i$ includes all the solution vector positions whose corresponding value is $i$ (up to here, there can be duplicates). If $sizeof(idx_i) > 1$ for some $i$, $idx_i$ is shuffled and all $s_j$ such that $j \in idx_i$ and $idx_i.indexof(j) > 0$ are re-set to -1. Basically, for each $idx_i$, the value $i$ is kept by the $s_j$ corresponding to the first index after the shuffling. By doing this, duplicate nodes in the cycle are removed. \\
Once all values in $s$ have been made unique or equal to $-1$, let us define $L = A \setminus K$. In detail, $L$ contains all the nodes that have not yet been assigned to a position in the cycle. Then, for each $i$ in $\{0, ..., n-1\}$ such that $s_i = -1$, a random value $l \in L$ is chosen and the set of remaining nodes is updated accordingly, i.e., $s_i = l$ and $L = L \setminus l$. At the end of this procedure, all the empty positions have an assigned value, i.e., $s_i \neq -1\;\ \forall\; i \in \{0, ..., n-1\}$, and the solution is valid.

The refinement procedure does not affect the original solution if it is already valid; if the solution is not valid, the algorithm does not affect the nodes whose position is already unique in the cycle.

\paragraph{TSP results discussion}
Starting from the smallest problem (with $n$ = 10), QALS has performed better than Embedding Composite but not as good as Hybrid, which has found a solution that is very close to the global minimum, or the Brute Force approach, which has (obviously) found the global minimum solution. 
Instead, with $n$ = 12, Embedding Composite has turned out to be better. Regarding $n$ = 14, which is the biggest dimension that Embedding Composite is able to manage, the approach just mentioned has produced again better solutions w.r.t. QALS. As concerns $n$ between 32 and 72, only QALS and Hybrid work, with Hybrid being always faster and better than QALS. Finally, only Hybrid supports problem with $n$ = 74. A careful reader might spot that this should not be possible because the size of the QUBO problem exceeds the number of available qubits. Probably, if a graph is not complete (none of ours is complete), Hybrid ignores the non-connected edges saving a lot of unused qubits. It is also worth remarking that Hybrid's good results are mostly related to the nature of the method, which runs multiple solvers in parallel. Eventually, the performance of both Embedding Composite and QALS could probably improve by considering a far higher number of annealer measurements ($k \gg 5$). Indeed, as explained in the \textit{NPP results discussion} paragraph, the annealer is subjected to noise and temperature effects, and larger output samples would increase the probability of obtaining a solution close to the optimum. Nevertheless, the time available to us was, again, too little to consider higher $k$ values.

\section{Conclusion}
\label{sec:conclusion}
\noindent
In this paper we presented two implementations of Quantum Annealing Learning Search, a quantum-classical algorithm proposed recently for solving QUBO problems, and their tests on two optimisation problems (Number Partitioning Problem and Travelling Salesman Problem). The original proposal of QALS included an argument on its asymptotic convergence and promised to treat QUBO problems that are not directly mappable to the machine topology. This research provides the empirical testing needed in order to assess QALS's practical applicability. The two implementations are in  C++ and Python, respectively and they run on the D-Wave machine. The two problems addressed in the tests have complementary characteristics: NPP is naturally representable as a QUBO problem, whereas TSP requires the setting of penalties to encode the constraints. As a baseline, we compared the results with the Hybrid procedure available in the D-wave suite.

The implementation in C++ has the advantage of having better performance in the classical part but pays the price of the coupling with the D-Wave APIs, which are currently available in Python only; for this very reason, the Python implementation proves to be competitive. The development of both codes required specific implementation choices with respect to the original pseudocode of QALS. In particular, the implementations include more efficient data structures and specific choices for the classical pseudorandom number generation in C++. We argue that the availability of C++ interfaces to the D-Wave machines, once present and now apparently discontinued, would make the development of efficient hybrid algorithms easier. 

The results obtained on NPP show that quantum annealing, in general, and QALS, in particular, are not of any practical use in this case. In fact, the classical exact algorithm used for NPP outperforms both QALS and Hybrid, and the problem can be considered practically solved by the available classical exact solution. In our opinion, NPP, despite its natural representability as a QUBO problem, is not the kind of problem on which we can expect an advantage using the quantum architecture. Hybrid includes  classical methods, so we argue that it works well on this particular problem because it exploits the classical efficiency.

Instead, the application to TSP shows the practical potential of QALS. The results show that QALS is able to process problems with a higher size than the Embedding Composite procedure used to map the problem to the topology; this shows that the goal of the introduction of QALS, namely treating QUBO problems not directly mappable to the machine topology, has been empirically fulfilled. However, both QALS and EC do not directly provide a valid solution; indeed, the solution has to be manipulated in order to become valid for the problem. Again, Hybrid takes full advantage of the races with classical methods, reaching good results.

We conclude that our implementations of QALS work on TSP and could have practical potential on hard problems where the QUBO representation is not directly mappable to the topology. We note that the comparison with respect to Hybrid is a necessary baseline but, given its nature of race among different algorithms, cannot be an effective competitor for drawing scientific conclusions. Both implementations are available under the GPLv2 licence \cite{c++git,pythongit}.

\section*{Acknowledgements}
\noindent
This work was supported by Q@TN, the joint lab between University of Trento, FBK-Fondazione Bruno Kessler, INFN-National Institute for Nuclear Physics and CNR-National Research Council. In addition, the authors gratefully acknowledge the J\"ulich Supercomputing Center (\url{https://www.fz-juelich.de/ias/jsc}) for funding this project by providing computing time through the J\"ulich UNified Infrastructure of Quantum computing (JUNIQ) on the D-Wave quantum annealer.


\begin{thebibliography}{000}

\bibitem{PhysRevE.58.5355}
T. Kadowaki and H. Nishimori (Nov 1998), {\it Quantum annealing in the transverse Ising model}, Phys. Rev. E, Vol.58, pp. 5355-5363.

\bibitem{bian2020solving}
Z. Bian and F. Chudak and W. Macready and A. Roy and R. Sebastiani and S. Varotti (2020), {\it Solving SAT (and MaxSAT) with a quantum annealer: Foundations, encodings, and preliminary results}, Information and Computation, Vol.275, pp. 104609.

\bibitem{preskill2018quantum}
J. Preskill (2018), {\it Quantum Computing in the NISQ era and beyond}, Quantum, Vol.2, pp. 79.

\bibitem{mcclean2016theory}
J. R. McClean and J. Romero and R. Babbush and A. Aspuru-Guzik (2016), {\it The theory of variational hybrid quantum-classical algorithms}, New Journal of Physics, Vol.18., Num.2, pp. 023023.

\bibitem{ayanzadeh2020reinforcement}
R. Ayanzadeh and M. Halem and T. Finin (2020), {\it Reinforcement quantum annealing: A hybrid quantum learning automata}, Scientific Reports, Vol.10, Num.1, pp. 1-11.

\bibitem{QALS}
D. Pastorello and E. Blanzieri (Aug 2019), {\it Quantum annealing learning search for solving QUBO problems}, Quantum Information Processing, Vol.18, Num.10, pp. 303.

\bibitem{pastorello_learning_2021}
D. Pastorello and E. Blanzieri and V. Cavecchia (2021), {\it Learning adiabatic quantum algorithms over optimization problems}, Quantum Machine Intelligence, Vol.3., Num.1, pp. 2.

\bibitem{QUBOsurvey}
G. Kochenberger and J. Hao and F. Glover and M. Lewis and Z. L{\"u} and H. Wang and Y. Wang (2014), {\it The unconstrained binary quadratic programming problem: A survey}, Journal of Combinatorial Optimization, Vol.28, Num.1, pp. 58-81.

\bibitem{minor_embedding}
D-Wave Systems Inc., {\it Minor embedding}, \url{https://docs.dwavesys.com/docs/latest/c_gs_3.html#minor-embedding} [online, last access on 20 October 2021].

\bibitem{minor_embedding_example}
D-Wave Systems Inc., {\it Minor embedding example}, \url{https://docs.dwavesys.com/docs/latest/c_gs_7.html#getting-started-embedding} [online, last access on 20 October 2021].

\bibitem{pegasus}
D-Wave Systems Inc., {\it Pegasus topology}, \url{https://docs.dwavesys.com/docs/latest/c_gs_4.html#pegasus-graph} [online, last access on 24 February 2021].

\bibitem{embedding_composite}
D-Wave Systems Inc., {\it Embedding Composite},
\url{https://docs.ocean.dwavesys.com/en/stable/docs_system/reference/composites.html#embeddingcomposite} [online, last access on 20 October 2021].

\bibitem{leap_hybrid_sampler}
D-Wave Systems Inc., {\it LeapHybridSampler}, \url{https://docs.ocean.dwavesys.com/en/stable/docs_system/reference/samplers.html#dwave.system.samplers.LeapHybridSampler} [online, last access on 20 October 2021].

\bibitem{solver_params}
D-Wave Systems Inc., {\it D-Wave solver parameters}, \url{https://docs.dwavesys.com/docs/latest/c_solver_parameters.html} [online, last access on 20 October 2021].

\bibitem{solver_properties}
D-Wave Systems Inc., {\it D-Wave solver properties}, \url{https://docs.dwavesys.com/docs/latest/c_solver_properties.html} [online, last access on 20 October 2021].

\bibitem{embed}
J. Du (Oct 2005), {\it Embedding Python in C/C++}, \url{https://www.codeproject.com/Articles/11805/Embedding-Python-in-C-C-Part-I} [online, last access on 24 February 2021].

\bibitem{rand}
Stack Overflow (Mar 2016), {\it C++ rand()}, \url{https://stackoverflow.com/questions/10984974/why-do-people-say-there-is-modulo-bias-when-using-a-random-number-generator} [online, last access on 24 February 2021].

\bibitem{coins}
Ridgeback Network Defense Inc. (October 2018), {\it D-Wave random coin flip}, \url{https://github.com/ridgebacknet/dwave-tutorials/blob/master/fun/fun-coin.py} [online, last access on 26 February 2021].

\bibitem{mt}
M. Route (2017), {\it Radio-flaring Ultracool Dwarf Population Synthesis}, The Astrophysical Journal, Vol.845, Num.1, pp. 66.

\bibitem{glover2019tutorial}
F. Glover and G. Kochenberger and Y. Du (2019), {\it A Tutorial on Formulating and Using QUBO
Models}, 1811.11538.

\bibitem{kk}
N. Karmarker and R. M. Karp (1983), {\it The Differencing Method of Set Partitioning}, Technical report, EECS Department, University of California, Berkeley (USA).

\bibitem{Korf98}
R. E. Korf (1998), {\it A complete anytime algorithm for number partitioning}, Artificial Intelligence, Vol.106, Num.2, pp. 181-203.

\bibitem{Pedroso2013}
J. P. Pedroso and M. Kubo (2010), {\it Heuristics and exact methods for number partitioning}, European Journal of Operational Research, Vol.202, Num.1, pp. 73-81.

\bibitem{Lucas_2014}
A. Lucas (2014), {\it Ising formulations of many NP problems}, Frontiers in Physics, Vol.2.

\bibitem{tsp_implementation}
M. {St\c ech\l y} and P. Korponai\'c (Mar 2018), {\it Quantum TSP}, \url{https://github.com/BOHRTECHNOLOGY/quantum_tsp} [online, last access on 27 June 2021].

\bibitem{c++git}
T. De Min (2021), {\it C++ Quantum Annealing Learning Search}, \url{https://github.com/tdemin16/QALS-cpp}.

\bibitem{pythongit}
A. Bonomi (2021), {\it Python Quantum Annealing Learning Search}, \url{https://github.com/bonom/Quantum-Annealing-for-solving-QUBO-Problems}.

\end{thebibliography}
\end{document}